\documentclass[aps,prd,showpacs,floatfix,preprint]{revtex4}
\usepackage{amsmath,bm}
\usepackage{graphicx}
\usepackage{epsfig}
\begin{document}
%%%%%%%%%%%%%%%%%%%%%%%%%%%%%%%%%%%%%%%%%%%%%%%%%%%%%%%%%%%%%%%%%%%%
\title{Critical temperature of antikaon condensation in nuclear matter} 
\author{Sarmistha Banik$^{\rm (a,c)}$, Walter Greiner$^{\rm (a)}$ and 
Debades Bandyopadhyay$^{\rm (b)}$} 
\affiliation{$^{\rm (a)}$Frankfurt Institute for Advanced Studies (FIAS), J. W. Goethe Universit\"at, 
Ruth Moufang Strasse 1, D-60438 Frankfurt am Main, Germany}

\affiliation{$^{\rm (b)}$Theory Division and Centre for Astroparticle Physics, 
Saha Institute of Nuclear Physics, 1/AF Bidhannagar, Kolkata 700 064, India}

\affiliation{$^{\rm (c)}$ Variable Energy Cyclotron Centre (VECC), 1/AF 
Bidhannagar, Kolkata 700 064, India}

\begin{abstract}
We investigate the critical temperature of Bose-Einstein condensation of $K^-$ 
mesons in neutron star matter. This is studied within the framework of 
relativistic field theoretical models at finite temperature where 
nucleon-nucleon and (anti)kaon-nucleon interactions are mediated by the 
exchange of mesons. The melting of the antikaon condensate is studied for
different values of antikaon optical potential depths. We find that the 
critical temperature of antikaon condensation increases with baryon number 
density. Further it is noted that the critical temperature is lowered as  
antikaon optical potential becomes less attractive. We also construct the 
phase diagram of neutron star matter with $K^-$ condensate.

\end{abstract}

\maketitle

\section{Introduction}

In the pioneering work of Kaplan and Nelson \cite{Kap}, it was demonstrated
within the $SU(3)_L\times SU(3)_R$ chiral perturbation theory 
that antikaon ($K^-$ meson) condensation might be possible  
in dense baryonic matter formed in heavy ion collisions as well as in neutron 
stars. The underlying idea is that the Bose-Einstein condensation of 
$K^-$ mesons is driven by the   strongly attractive $K^-$-nucleon 
interaction. Consequently, the attractive antikaon-nucleon interaction
reduces the effective mass ($m_K^*$) and in-medium energy ($\omega_{K^-}$) 
of $K^-$ mesons. The $s$-wave $K^-$ condensation sets in when  
$\omega_{K^-}$ equals to the $K^-$ chemical potential $\mu_{K^-}$ 
which, in turn, is equal to the electron chemical potential $\mu_e$ for  
neutron (neutrino-free) star matter in $\beta$ equilibrium. The 
threshold density for $K^-$ condensation which sensitively depends on 
the nuclear equation of state (EoS) and the strength of the attractive 
antikaon optical potential depth, is about $(2-4)n_0$, where
$n_0$ is the normal nuclear matter density. 

There was considerable interest in the study of the properties
of kaons ($K$) and antikaons ($\bar K$) in dense nuclear matter as well as 
neutron star matter after the work of Kaplan and Nelson. Antikaon condensation 
in neutron star interior was studied in great details in the chiral 
perturbation theory \cite{Bro92,Tho,Ell,Lee,Pra97} as well as meson exchange 
model \cite{Gle99,Mut,Kno,Sch,Li,Pal,Bani1,Bani2,Bani3,Bani4}. 
The net effect of $K^-$ condensation in neutron star matter is 
to soften the EoS leading to a smaller maximum mass neutron star than that 
of the case without the condensate.

There is a growing interplay between the physics of dense matter  formed in
heavy ion collisions and the physics of neutron stars. The study of dense 
matter in future experiments at FAIR in GSI might reveal many new and 
interesting results. It would be 
possible to produce matter with baryon density a few times normal nuclear
matter density and temperature a few tens of MeV at FAIR. Under these 
circumstances, antikaon condensation might occur in dense matter as it was 
predicted by Nelson and Kaplan \cite{Kap}. In this case one has to investigate
antikaon condensation at finite temperature.

Antikaon condensation at finite temperature was already studied in connection
with neutron \cite{Muto} and protoneutron (newly born and neutrino trapped) 
stars \cite{Pons}. These studies were either related to dynamical
evolution of the condensation or metastability of protoneutron stars. However,
the melting of antikaon condensate in nuclear matter is so far not looked 
into. This motivates us to investigate the critical temperature of $K^-$ 
condensation in neutron star matter. Further, we wish to construct a phase 
diagram of nuclear matter involving $K^-$ condensate.

The organisation of the paper is the following. We discuss the composition and
EoS involving $K^-$ condensate at finite temperature in Section II. Results are
discussed in Section III and Summary is given in Section IV.

\section{Composition and Equation of State}
We consider a second order phase transition from hadronic
to $K^-$ condensed matter in neutron stars where both
the hadronic and $K^-$ condensed matter are described within the
framework of relativistic field theoretical models. Constituents
of matter are neutrons ($n$), protons ($p$), electrons, muons in both phases 
and also (anti)kaons in the condensed phase. The baryon-baryon interaction is 
mediated by the exchange of $\sigma$, $\omega$ and $\rho$ mesons. 
Both phases maintain  local charge neutrality and  beta-equilibrium conditions.
The baryon-baryon interaction is described by the following Lagrangian density 
\begin{eqnarray}
{\cal L}_B &=& \sum_{B=n,p} \bar\Psi_{B}\left(i\gamma_\mu{\partial^\mu} - m_B
+ g_{\sigma B} \sigma - g_{\omega B} \gamma_\mu \omega^\mu
- g_{\rho B}
\gamma_\mu{\mbox{\boldmath t}}_B \cdot
{\mbox{\boldmath $\rho$}}^\mu \right)\Psi_B\nonumber\\
&& + \frac{1}{2}\left( \partial_\mu \sigma\partial^\mu \sigma
- m_\sigma^2 \sigma^2\right) - U(\sigma) \nonumber\\
&& -\frac{1}{4} \omega_{\mu\nu}\omega^{\mu\nu}
+\frac{1}{2}m_\omega^2 \omega_\mu \omega^\mu
- \frac{1}{4}{\mbox {\boldmath $\rho$}}_{\mu\nu} \cdot
{\mbox {\boldmath $\rho$}}^{\mu\nu}
+ \frac{1}{2}m_\rho^2 {\mbox {\boldmath $\rho$}}_\mu \cdot
{\mbox {\boldmath $\rho$}}^\mu ~.
\end{eqnarray}
Here $\psi_B$ denotes the Dirac bispinor for baryons $B$ with vacuum mass $m_B$
and the isospin operator is ${\mbox {\boldmath t}}_B$. The scalar
self-interaction term \cite{Bog} is
\begin{equation}
U(\sigma) = \frac{1}{3} g_2 \sigma^3 + \frac{1}{4} g_3 \sigma^4 ~.
\end{equation}

The thermodynamic potential per unit volume for nucleons is given by 
\cite{Ser},
\begin{eqnarray}
\frac{\Omega_N}{V} &=& \frac{1}{2}m_\sigma^2 \sigma^2
+ \frac{1}{3} g_2 \sigma^3 + \frac{1}{4} g_3 \sigma^4  
- \frac{1}{2} m_\omega^2 \omega_0^2 
- \frac{1}{2} m_\rho^2 \rho_{03}^2  \nonumber \\
&& - 2T \sum_{i=n,p} \int \frac{d^3 k}{(2\pi)^3} 
[ln(1 + e^{-\beta(E^* - \nu_i)}) +
ln(1 + e^{-\beta(E^* + \nu_i)})] ~.  
\end{eqnarray}
Here the temperature is defined by $\beta = 1/T$ and 
$E^* = \sqrt{(k^2 + m_N^{*2})}$.
The effective baryon mass is $m_N^*=m_N - g_{\sigma N}\sigma$.
Neutron and proton chemical potentials are given by
$\mu_{n} = \nu_n + g_{\omega N}
\omega_0 - \frac {1}{2} g_{\rho N} \rho_{03}$ and
$\mu_{p} = \nu_p + g_{\omega N}
\omega_0 + \frac {1}{2} g_{\rho N} \rho_{03}$.
The number density of i(=n,p)th baryon is 
$n_i = 2 \int \frac{d^3k}{(2\pi)^3} \left({\frac{1}{e^{\beta(E^*-\nu_i)} 
+ 1}} - {\frac{1}{e^{\beta(E^*+\nu_i)} + 1}}\right)$. The pressure due to
nucleons is given by $P_N = - {\Omega_N}/V$. The explicit form of the energy 
density is given below,
\begin{eqnarray}
\epsilon_N &=& \frac{1}{2}m_\sigma^2 \sigma^2
+ \frac{1}{3} g_2 \sigma^3 + \frac{1}{4} g_3 \sigma^4  
+ \frac{1}{2} m_\omega^2 \omega_0^2 
+ \frac{1}{2} m_\rho^2 \rho_{03}^2  \nonumber \\
&& + 2 \sum_{i=n,p} \int \frac{d^3 k}{(2\pi)^3} E^* 
\left({\frac{1}{e^{\beta(E^*-\nu_i)} 
+ 1}} + {\frac{1}{e^{\beta(E^*+\nu_i)} + 1}}\right)~.  
\end{eqnarray}
We can calculate the entropy density of nucleons using 
${\cal S}_N = \beta \left(\epsilon_N + P_N - \sum_{i=n,p} \mu_i n_i \right)$. 
The entropy density per baryon is given by ${{\cal S}_N}/{n_b}$ where $n_b$ is
the total baryon density.
Similarly, we calculate number densities, energy densities and pressures of
electrons, muons and their antiparticles using the thermodynamic potential
per unit volume
\begin{equation}
\frac{\Omega_L}{V} = - 2T \sum_l \int \frac{d^3 k}{(2\pi)^3} [ ln(1 + 
e^{-\beta(E_l - \mu_l)})
+ ln(1 + e^{-\beta(E_l + \mu_l)})]~.
\end{equation}

We treat the (anti)kaon-baryon interaction in the same footing as the
baryon-baryon interaction. The Lagrangian density for (anti)kaons in the
minimal coupling scheme is \cite{Gle99,Bani2},
\begin{equation}
{\cal L}_K = D^*_\mu{\bar K} D^\mu K - m_K^{* 2} {\bar K} K ~,
\end{equation}
where the covariant derivative is
$D_\mu = \partial_\mu + ig_{\omega K}{\omega_\mu} 
+ i g_{\rho K}
{\mbox{\boldmath t}}_K \cdot {\mbox{\boldmath $\rho$}}_\mu$ and
the effective mass of (anti)kaons is
$m_K^* = m_K - g_{\sigma K} \sigma$.

We adopt the finite temperature treatment of antikaon condensation by Pons et
al. \cite{Pons} in our calculation. Once the antikaon condensation sets in, 
the thermodynamic potential for antikaons is given by,
\begin{equation}
\frac {\Omega_K}{V} = T \int \frac{d^3p}{(2\pi)^3} [ ln(1 - 
e^{-\beta(\omega_{K^-} - \mu)}) + 
 ln(1 - e^{-\beta(\omega_{K^+} + \mu)})]~.
\end{equation}
The in-medium energies of $K^{\pm}$ mesons are given by
\begin{equation}
\omega_{K^{\pm}} =  \sqrt {(p^2 + m_K^{*2})} \pm \left( g_{\omega K} \omega_0
+ \frac {1}{2} g_{\rho K} \rho_{03} \right)~,
\end{equation}
and $\mu$ is the chemical potential of $K^-$ mesons and is given by 
$\mu = \mu_n -\mu_p$. The threshold condition for $K^-$ condensation is given
by
$\mu = \omega_{K^{-}} =   m_K^* - g_{\omega K} \omega_0
- \frac {1}{2} g_{\rho K} \rho_{03}$~.

The meson field equations in the presence of the condensed are, 
\begin{eqnarray}
m_\sigma^2\sigma &=& -\frac{\partial U}{\partial\sigma}
+ \sum_{B=n,p} g_{\sigma N} n_B^S
+ g_{\sigma K} \left(n_{K}^C + n_{K}^S \right)~,\\
m_\omega^2\omega_0 &=& \sum_{B=n,p} g_{\omega N} n_B
- g_{\omega K}  n_K ~,\\
m_\rho^2\rho_{03} &=& \sum_{B=n,p} g_{\rho N} I_{3B} n_B
- \frac {1}{2} g_{\rho K} n_K ~,
\end{eqnarray}
where isospin projection for baryons $B$ is $I_{3B}$.

Scalar densities for baryons and (anti)kaons are respectively
\begin{eqnarray}
n_B^S &=& 
2 \int \frac{d^3 k}{(2\pi)^3} \frac{m_B^*}{E^*} 
\left({\frac{1}{e^{\beta(E^*-\nu_B)} 
+ 1}} + {\frac{1}{e^{\beta(E^*+\nu_B)} + 1}}\right) ~,\\
n_K^S &=&  
\int \frac{d^3 p}{(2\pi)^3} \frac{m_K^*}{\sqrt{p^2 + m_K^{*2}}} 
\left({\frac{1}{e^{\beta(\omega_{K^-}-\mu)} 
- 1}} + {\frac{1}{e^{\beta(\omega_{K^+}+\mu)} - 1}}\right)~.  
\end{eqnarray}

The net (anti)kaon number density is given by
\begin{equation}
n_K = n_K^C + n_K^{T}~,
\end{equation}
where $n_K^C$ gives the condensate density and $n_K^{T}$ represents the thermal
density. Again the thermal (anti)kaon density is given by,
\begin{equation}
n_K^{T} =  
\int \frac{d^3 p}{(2\pi)^3} 
\left({\frac{1}{e^{\beta(\omega_{K^-}-\mu)} 
- 1}} - {\frac{1}{e^{\beta(\omega_{K^+}+\mu)} - 1}}\right)~.  
\end{equation}

It is straight forward to calculate the pressure of thermal (anti)kaons 
$P_K = -{\Omega_K}/{V}$. The condensate does not contribute to the pressure.
The energy density of (anti)kaons is given by 
\begin{equation}
\epsilon_K = m_K^* n_K^C + \left( g_{\omega K} \omega_0
+ \frac {1}{2} g_{\rho K} \rho_{03} \right) n_K^T
+
\int \frac{d^3 p}{(2\pi)^3} 
\left({\frac{\omega_{K^-}}{e^{\beta(\omega_{K^-}-\mu)} 
- 1}} + {\frac{\omega_{K^+}}{e^{\beta(\omega_{K^+}+\mu)} - 1}}\right)~.  
\end{equation}

The first term is the contribution due to $K^-$ condensate and second and third
terms are the thermal contributions to the energy density.
The total energy density in the condensed phase is 
$\epsilon = \epsilon_N + \epsilon_K + \epsilon_l$. The entropy density of 
(anti)kaons is 
given by ${\cal S}_K = \beta \left(\epsilon_K + P_K -  \mu n_K \right)$. 
The total entropy per baryon is given by 
$S = ({\cal S}_N + {\cal S}_K + {\cal S}_l)/n_b$, where ${\cal S}_l$ is the
entropy density of leptons \cite{Pons}. 

It is to be noted that for $s$-wave ${\bar K}$ condensation at T=0, the scalar 
and vector densities of antikaons are same and those are given by \cite{Gle99}
\begin{equation}
n^C_{K} = 2\left( \omega_{K^-} + g_{\omega K} \omega_0
+ \frac{1}{2} g_{\rho K} \rho_{03} \right) {\bar K} K
= 2m^*_K {\bar K} K  ~.
\end{equation}

Further we impose the charge neutrality and $\beta$ equilibrium conditions 
which are given by,
\begin{eqnarray}
n_p - n_K - n_e - n_{\mu}=0~,\\
\mu_n - \mu_p = \mu_e~.
\end{eqnarray}

\section{Results and Discussion}

Nucleon-meson coupling constants are calculated by reproducing the nuclear 
matter saturation properties such as binding energy -16 MeV, saturation 
density ($n_0$) 0.153 $fm^{-3}$, asymmetry energy coefficient 32.5 MeV, 
effective nucleon mass ($m_N^*/m_n$) 0.70 and incompressibility $K = 300$ MeV. 
This parameter set is known as GM1 set \cite{Gle91}. We
have taken these nucleon-meson couplings from the Table I of Ref. \cite{Bani2}.

The kaon-scalar meson coupling is determined from the real part of antikaon
optical potential depth at normal nuclear matter density

\begin{equation}
U_K(n_0) = -g_{\sigma K} \sigma - g_{\omega K} \omega_0~.
\end{equation}
Kaon-vector meson coupling are obtained from the quark model and isospin 
counting rules. They are given by,

\begin{equation}
g_{\omega K} = \frac {1}{3} g_{\omega N}, ~ g_{\rho K} = g_{\rho N}~.
\end{equation}

The value of antikaon optical potential depth is a debatable issue.
The analysis of kaonic atom data yielded the real part of antikaon optical
potential depth $U_K = -180 \pm 20$ MeV at normal nuclear matter density 
\cite{Fri94,Fri99,Gal07}. On the other hand, the chiral model suggests the
strength of the optical potential $\sim -60$ MeV \cite{Tol}. However, the
$K^-$ condensation sets in when the magnitude of antikaon optical potential 
depth is $\sim 100$ MeV or more. We perform this calculation for 
$U_K = -100, -160$ MeV and the corresponding kaon-scalar coupling constants 
are taken from Table II of Ref. \cite{Bani2}. 

We study the evolution of a hot neutron star after the emission of trapped 
neutrinos to the cold neutron star in this calculation. As the temperature
varies from the center to the surface, we consider certain fixed entropy per
baryon situations corresponding to the hot neutron star. Here we consider 
second order antikaon condensation for both values of  $U_K = -100, -160$ MeV. 
It was already noted by Pons et al. \cite{Pons} that the phase transition which
was first order at zero temperature for moderate values of antikaon optical 
potential depth,
became second order phase transition at finite temperature. Further they 
observed that when the antikaon condensation was a first order phase 
transition at finite temperature for strongly attractive antikaon potential, 
its pressure-energy density relation was similar to that of a second order 
phase transition. We show the temperature as a function of baryon density for 
$S=2$ and  $U_K = -100, -160$ MeV in Figure 1. Temperature increases with 
baryon density for both cases. The temperature for $U_K = -160$ falls below
that of $U_K = -100$ MeV case due to the early onset of $K^-$ condensation 
in the former case.  

At zero temperature, the threshold density of $K^-$ condensation is 
3.4$n_0$ for $U_K = -100$ MeV whereas it is 2.4$n_0$ for $U_K = -160$ MeV. 
Finite temperature effects shift the threshold of the condensation to higher
density \cite{Pons}. For $S=2$ and $U_K = -100, -160$ MeV, threshold
densities of $K^-$ condensation are 4.0$n_0$ and 2.7$n_0$ respectively. 
In Figure 2 the populations of thermal (anti)kaons (dotted line) and $K^-$ 
mesons in the condensate (dashed line) in $\beta$-equilibrated matter are shown
with baryon density for $S=2$ case and $U_K = -160$ MeV. The total density of 
(anti)kaons is given by the solid line. The thermal (anti)kaons are populated 
well before the threshold of the condensate. As soon as the condensation sets 
in, the density of $K^-$ mesons takes over the thermal contribution. It is
worth mentioning here that the onset of antikaon condensation for a 
particular value of $U_K$ is shifted to higher density for larger value of $S$.

Now we focus on the determination of critical temperature ($T_C$) for antikaon 
condensation. The condensate does not exist for temperatures $T > T_C$. The 
density of $K^-$ mesons in the condensate is a function of baryon density and 
temperature. We find that the density of antikaons in the condensate increases 
with baryon density and temperature as it is evident from Figures 1 and 2. The 
ratio of the density of $K^{-}$ mesons in the condensate ($n_K^C(T)$) at finite
temperature to that ($n_K^C(T=0)$) of zero temperature is plotted with 
temperature for several fixed baryon densities and 
$U_K = -160$ MeV in Figure 3. Each curve in the figure corresponds to a fixed
baryon density. In each case, the density of antikaons drops
with increasing temperature and the condensate melts down at certain 
temperature which is defined as the critical temperature of the condensation.
We note that the critical temperature increases as baryon density increases.
Similar investigation is done for $U_K = -100$ MeV and different fixed baryon 
densities. These results are shown in Figure 4. Like Fig. 3 we obtain the same 
feature of the critical temperature as a function of baryon density for 
$U_K = -100$ MeV. However, the comparison of same fixed baryon density curve,
for example 4$n_0$ in Figures 3 and 4, shows that the critical temperature for
$U_K = -100$ MeV is much smaller than that of $U_K = -160$ MeV case. It is to 
be noted here that the hot neutron star after trapped neutrinos are emitted has 
a maximum entropy per baryon $S=2$ \cite{Pons}. However, we have obtained large
values of temperatures for certain fixed baryon densities in Figures 3 
and 4 which correspond to $S>2$. Such high temperature or the corresponding 
entropy per baryon might not be relevant for a protoneutron star but
it could be important for dense matter formed in heavy ion experiments in 
upcoming accelerator facilities.

Entropy per baryon is plotted with temperature for $U_K = -160, -100$ MeV in
Fig. 5 and Fig. 6 respectively. Each curve in both figures corresponds to
a fixed value of baryon density. The end point of each curve in Fig. 5 and 
Fig. 6 indicates the corresponding critical temperature as it is obtained in 
Fig. 3 and 
Fig. 4 respectively. For $U_K = -160$ MeV, we vary entropy per baryon from 
$S =$ 0 to 5.5 to get critical temperatures whereas we use $S =$ 0 to 2.75 for 
the calculation with $U_K = -100$ MeV.

Now we know the critical temperature as a function of baryon density. It is 
straight forward to construct a phase diagram of nuclear matter with $K^-$
condensate. Figure 7 displays temperature versus baryon density for $\beta$
equilibrated nuclear matter with the antikaon condensate. The critical 
temperature lines for $U_K = -160$ MeV (solid line) and $U_K = -100$ MeV 
(dashed line) separate two phases. The condensed phase exists below the
critical temperature lines and the nuclear matter phase above them. Similarly
a phase diagram with antikaon condensate could be constructed for heavy ion 
collisions which might be probed in future experiments at FAIR, GSI. 
It is to be noted that the attractive antikaon-nucleon interaction drives
antikaon condensation in neutron stars as well as heavy ion collisions.
Heavy ion collisions in 
Compressed Baryon Matter (CBM) experiment in FAIR might produce matter with
density a few times normal nuclear matter density and temperature a few tens of
MeV. Such a scenario would be relevant for medium modification of 
the mass and energy of $K^-$ meson due to attractive nucleon-antikaon 
interaction \cite{Mis}. Consequently, it might lead to $K^-$ condensate at 
finite temperature. 
The formation of an antikaon condensate in heavy ion collisions might lead
to enhanced production of strange hadrons \cite{Kap}.

\section{summary}
We have investigated the critical temperature of $K^-$ condensation in 
$\beta$ equilibrated nuclear matter for antikaon optical
potential depths $U_K = -100, -160$ MeV within the framework of field 
theoretical models. Critical temperature of antikaon condensation increases with
baryon density. For the same baryon density, the critical temperature is 
larger in case of $U_K = -160$ MeV than that of $U_K = -100$ MeV. We 
construct the phase diagram of $\beta$ equilibrated nuclear matter with $K^-$ 
condensate. Further we discuss antikaon condensation in heavy ion collisions
and its implications to future heavy ion experiments at FAIR, GSI.

\section{Acknowledgement}
S.B. thanks the Alexander von Humboldt Foundation for the support. 

%%%%%%%%%%%%%%%%%%%%%%%%%%%%%%%%%%%%%%%%%%%%%%%%%%%%%%%%%%%%%%%%%%%%

%%%%%%%%%%%%%%%%%%%%%%%%%%%%%%%%%%%%%%%%%%%%%%%%%%%%%%%%%%%%%%%%%%%
\newpage 
\vspace{-2cm}

{\centerline{
\epsfxsize=12cm
\epsfysize=14cm
\epsffile{fig1.eps}
}}

\vspace{4.0cm}

\noindent{\small{
FIG. 1. Temperature is plotted with normalised baryon density for entropy
per baryon $S = 2$ and  antikaon optical potential depth at normal nuclear 
matter density $U_{\bar K} = -100, -160$ MeV.}}

\newpage
\vspace{-2cm}

{\centerline{
\epsfxsize=12cm
\epsfysize=14cm
\epsffile{fig2.eps}
}}

\vspace{4.0cm}

\noindent{\small{
FIG. 2. Kaon number densities $n_K$ in 
$\beta$-equilibrated nuclear matter including $K^-$ condensate are shown as
a function of normalised baryon density for entropy
per baryon $S = 2$ and  antikaon optical potential depth at normal nuclear 
matter density $U_{\bar K} = -160$ MeV.}}
\newpage
\vspace{-2cm}

{\centerline{
\epsfxsize=12cm
\epsfysize=14cm
\epsffile{fig3.eps}
}}

\vspace{4.0cm}

\noindent{\small{
FIG. 3. The ratio of the density of $K^-$ mesons in the condensate at a nonzero
temperature to that of zero temperature is plotted with temperature for 
fixed baryon number densities and antikaon optical potential depth at normal 
nuclear matter density $U_{\bar K} = -160$ MeV.}}

\newpage
\vspace{-2cm}

{\centerline{
\epsfxsize=12cm
\epsfysize=14cm
\epsffile{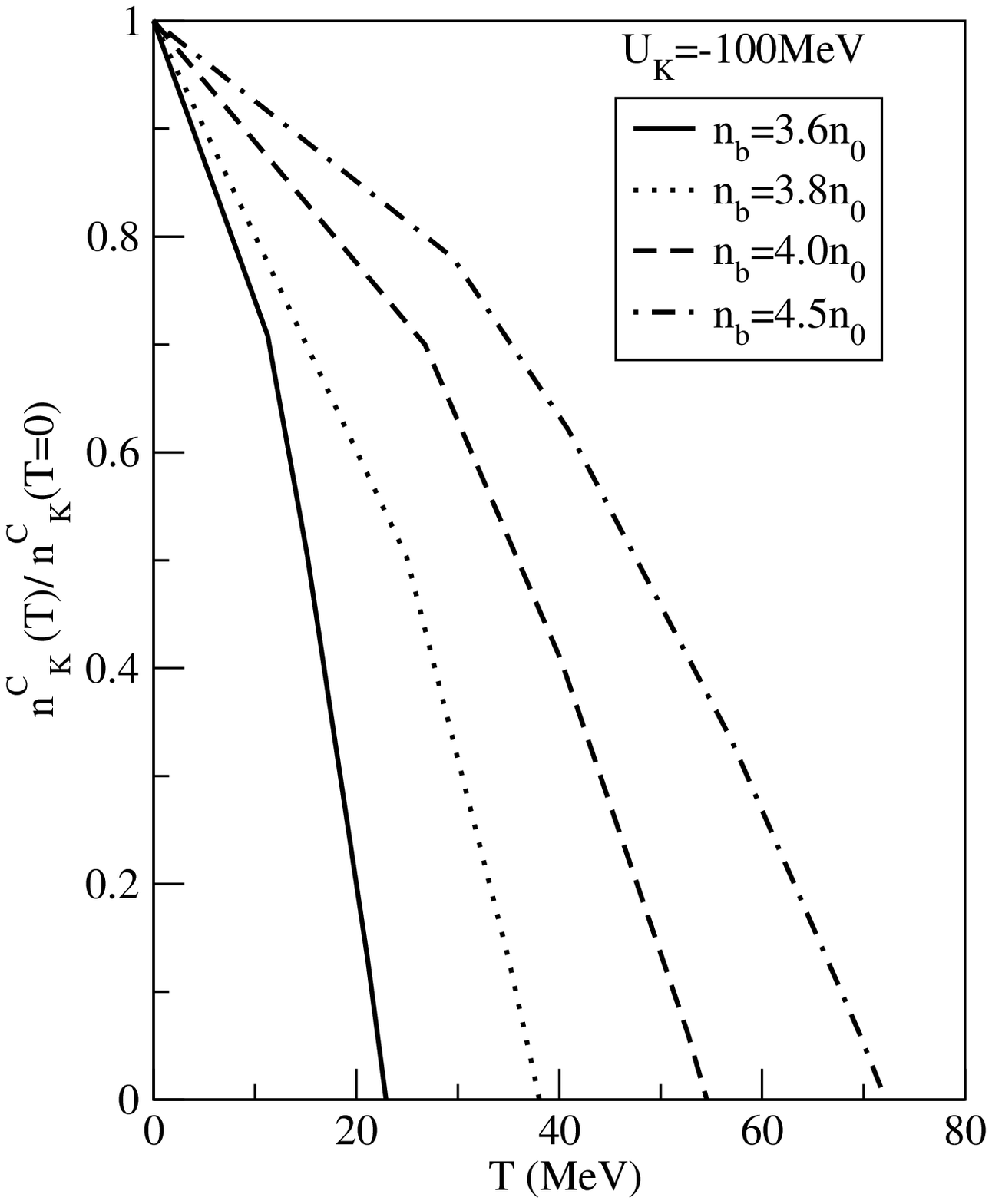}
}}

\vspace{4.0cm}

\noindent{\small{
FIG. 4. Same as Figure 3 but for  antikaon optical potential depth at normal 
nuclear matter density $U_{\bar K} = -100$ MeV.}}

\newpage
\vspace{-2cm}

{\centerline{
\epsfxsize=12cm
\epsfysize=14cm
\epsffile{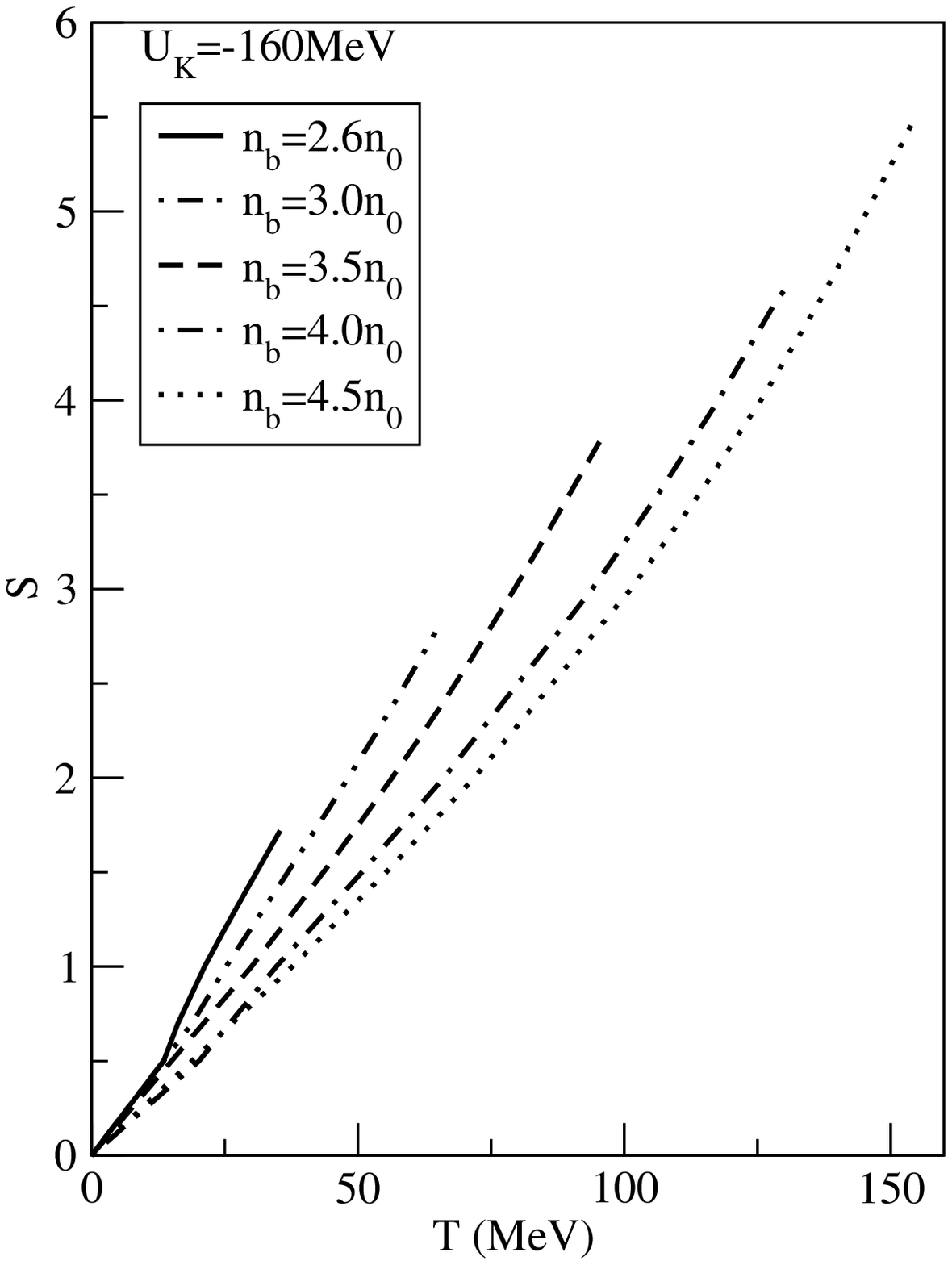}
}}

\vspace{4.0cm}

\noindent{\small{
FIG. 5. Entropy per baryon is plotted with temperature for 
fixed baryon number densities and antikaon optical potential depth at normal 
nuclear matter density $U_{\bar K} = -160$ MeV.}}

\newpage
\vspace{-2cm}

{\centerline{
\epsfxsize=12cm
\epsfysize=14cm
\epsffile{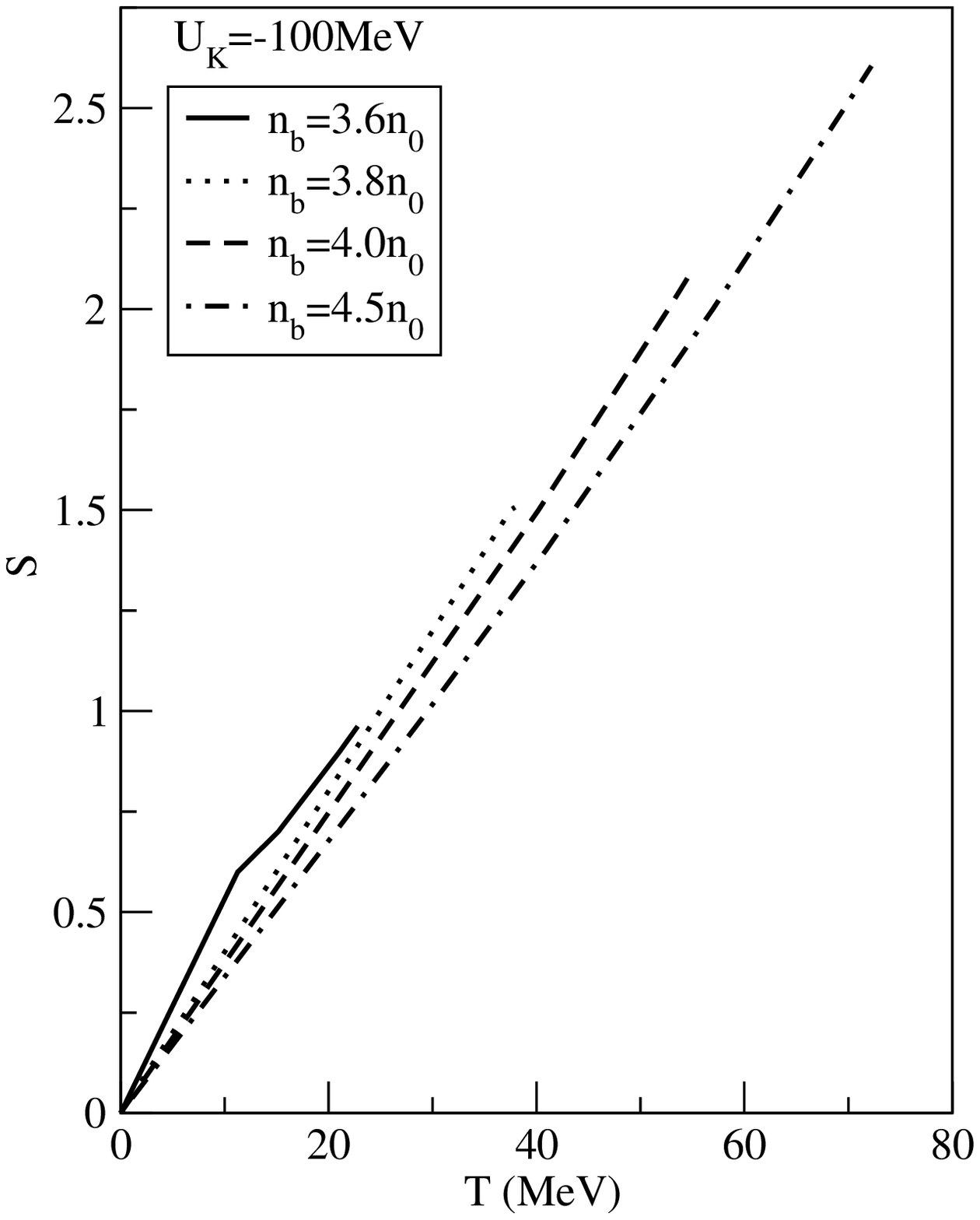}
}}

\vspace{4.0cm}

\noindent{\small{
FIG. 6. Same as Figure 5 but for  antikaon optical potential depth at normal 
nuclear matter density $U_{\bar K} = -100$ MeV.}}

\newpage
\vspace{-2cm}

{\centerline{
\epsfxsize=12cm
\epsfysize=14cm
\epsffile{fig7.eps}
}}

\vspace{4.0cm}

\noindent{\small{
FIG. 7. Phase diagram of nuclear matter with $K^-$ condensate. The solid and
dashed lines correspond to critical temperatures of $K^-$ condensation for 
antikaon optical potential depths at normal nuclear matter density 
$U_{\bar K} = -100, -160$ MeV, respectively.}}
\end{document}